\documentclass[11pt]{article}
\usepackage[]{amsmath}
\usepackage{setspace}
\usepackage{fullpage}
\usepackage[normalem]{ulem}
\usepackage{framed}
\usepackage{xcolor}
\usepackage{lineno}
\usepackage{hyperref}
\usepackage{makeidx}
\usepackage{graphicx}
\graphicspath{  {./Figs/}  } 
\usepackage[all]{xy}	
\usepackage[compress,authoryear]{natbib}

\newcommand{\unit}[1]{\ensuremath{\, \mathrm{#1}}}

\usepackage{mdwlist}
\newcommand{\RNum}[1]{\uppercase\expandafter{\romannumeral #1\relax}}
\author{Michel Crucifix \textsuperscript{1,2}, 
       Guillaume Lenoir\textsuperscript{1},  and
       Takahito Mitsui\textsuperscript{1}\\
        \textsuperscript{1} 
        \footnotesize{
        Universit\'e catholique de Louvain, 
        Georges Lema\^itre Centre for Earth and Climate Research, 
        Earth and Life Institute} \\ \footnotesize{ BE-1348 Louvain-la-Neuve, Belgium  } \\
        \textsuperscript{2} \footnotesize {Belgian National Fund of Scientific Research, 
        Rue d'Egmont, 5, BE-1000 Brussels, Belgium }}

        \title{Challenges for ice age dynamics: a~dynamical systems perspective \\ \medskip 
        \footnotesize{\textbf{submitted} as a chapter for Nonlinear and Stochastic Climate Dynamics, 
        edited by Christian L. E. Franzke (University of Hamburg, Germany) and Terence J. O'Kane (CSIRO, Australia), for publication by Cambridge University Press.  \footnote{Note : this version will differ from the final, accepted version of the published book, following the conditions given at \url{http://www.cambridge.org/gb/academic/cambridge-open-access/cambridge-open-access-books/green-archiving-policy-books}}. The final publication is expected late 2016. } }

\makeindex
\begin{document}
%\linenumbers
%\modulolinenumbers[5]
\maketitle

\doublespace
\begin{abstract}
This  chapter is dedicated to the slow dynamics \index{slow dynamics} of the climate system, at time scales of one~thousand to one million years. We focus specifically on the phenomenon of ice ages \index{ice ages}that has characterised the slow evolution of climate over the Quaternary. \index{Quaternary}
Ice ages are a form of variability featuring interactions between different large-scale components and processes in the climate system, including ice sheet, deep-ocean and carbon cycle dynamics. 
This  variability is also at least partly controlled  by changes in the seasonal and latitudinal incoming solar radiation associated with the combined effects of changes in Earth's orbit shape, precession of equinoxes, and changes in obliquity. A number of possible mechanisms are reviewed in this chapter. We stress that the nature of the interactions between these slow dynamics and faster modes of variability, such as millennium and centennial modes of variability, are still poorly understood. For example, whether the time sequence of ice ages is robustly determined or not by the astronomical forcing is a matter of debate.
These questions need to be addressed with  a range of models. We propose to use \index{stochastic parameterisation}stochastic parameterisations in the lower resolution models (focusing on climate time scales) to account for weather and macro-weather \index{macro-weather} dynamics simulated with higher resolution models. We discuss challenges---including statistical challenges---and possible methods associated with this programme. 
\end{abstract}

\newpage
\clearpage

\tableofcontents

\newpage
\clearpage

\newpage
\clearpage

\begin{flushright}
 If [\dots] we look in the 21st century and make an optimistic forecast on the type of computer which will be available [\dots] We may construct a~super-model [\ldots]  When we integrate the equations, if they are correct,  we shall necessarily obtain changes in climate, including the great ice ages. \emph{--- Edward Lorenz, 1970}\nocite{Lorenz70aa}

\end{flushright}

\section{The ice age phenomenon}

\subsection{Short summary of observational evidence \label{sect:ia}}
Glacial cycles, or ice ages, are a~form of climate variability that can be characterised as the succession of interglacial (similar to today) and glacial conditions over time scales of several tens of thousands of years. 
At the Last Glacial Maximum---21,000 years ago, henceforth noted 21~ka BP---the sea-level was 120~m lower than today. This water had accumulated in the form of ice on the continents, the majority of which was located in the Northern Hemisphere, in the current locations of Canada, Scandinavia~and the British Isles \citep{Clark02aa,peltier04ice5g}.

The recurrent character of glacial conditions was first identified from inspection of European alluvial terraces \citep{Penck09aa}.
It is nowadays documented from, among others, marine cores \citep{lisiecki05lr04, Elderfield12aa, Rohling14aa}, ice cores \citep{petit99, Kawamura07aa, EPICA04}, loess \citep{Guo00aa} and speleothems \citep{winograd92aa}. 

\begin{figure}
\includegraphics[scale=0.6]{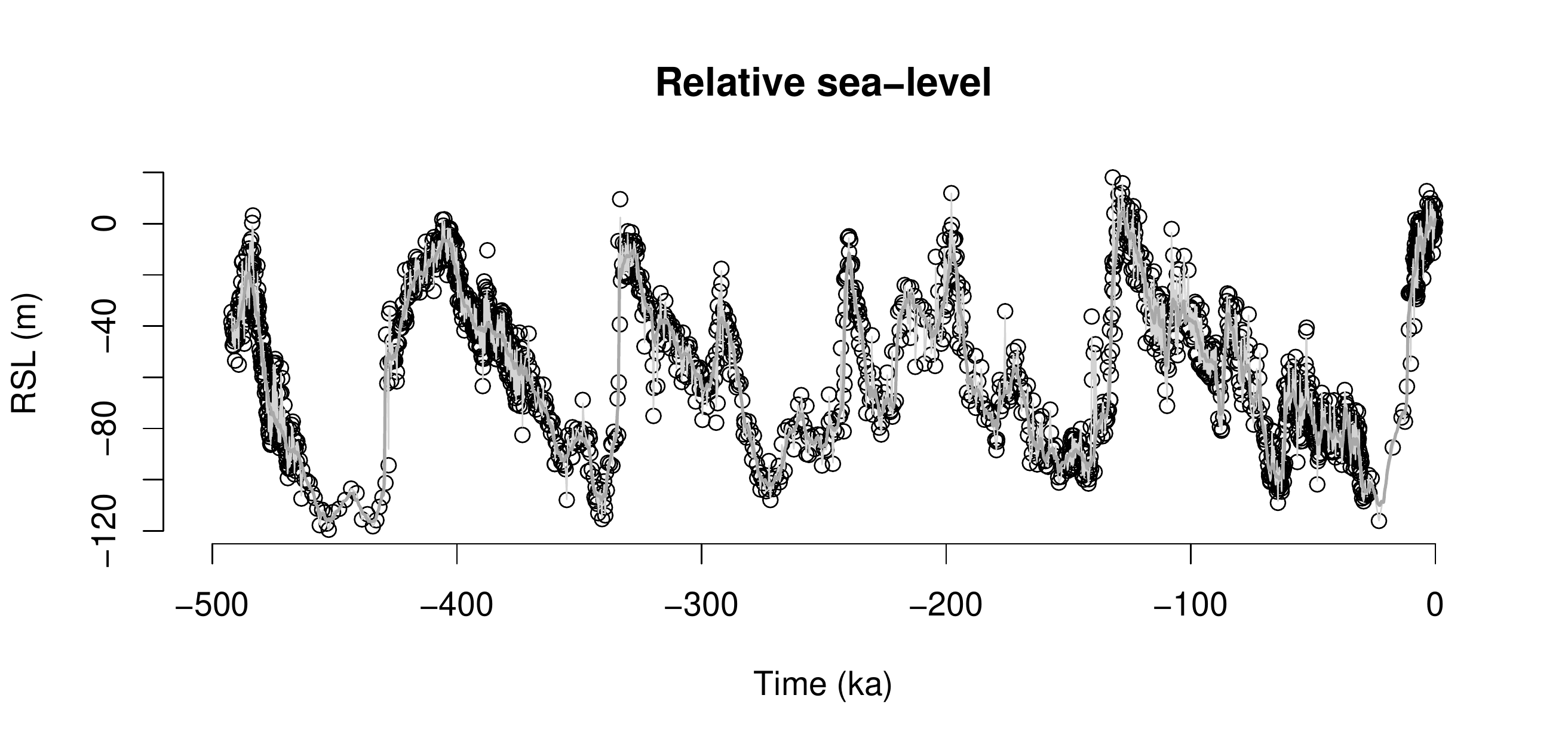}
\includegraphics[scale=0.6]{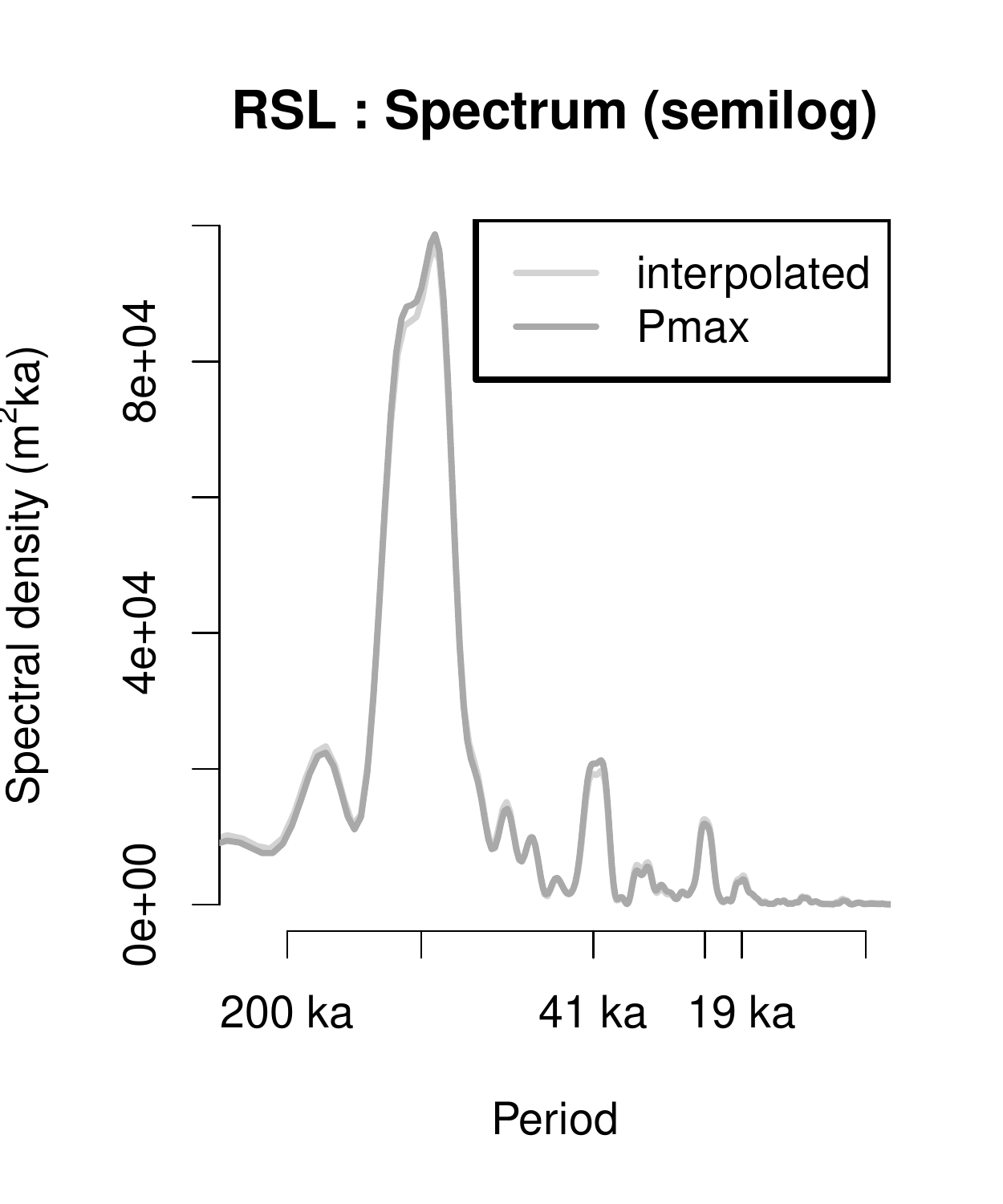}
\includegraphics[scale=0.6]{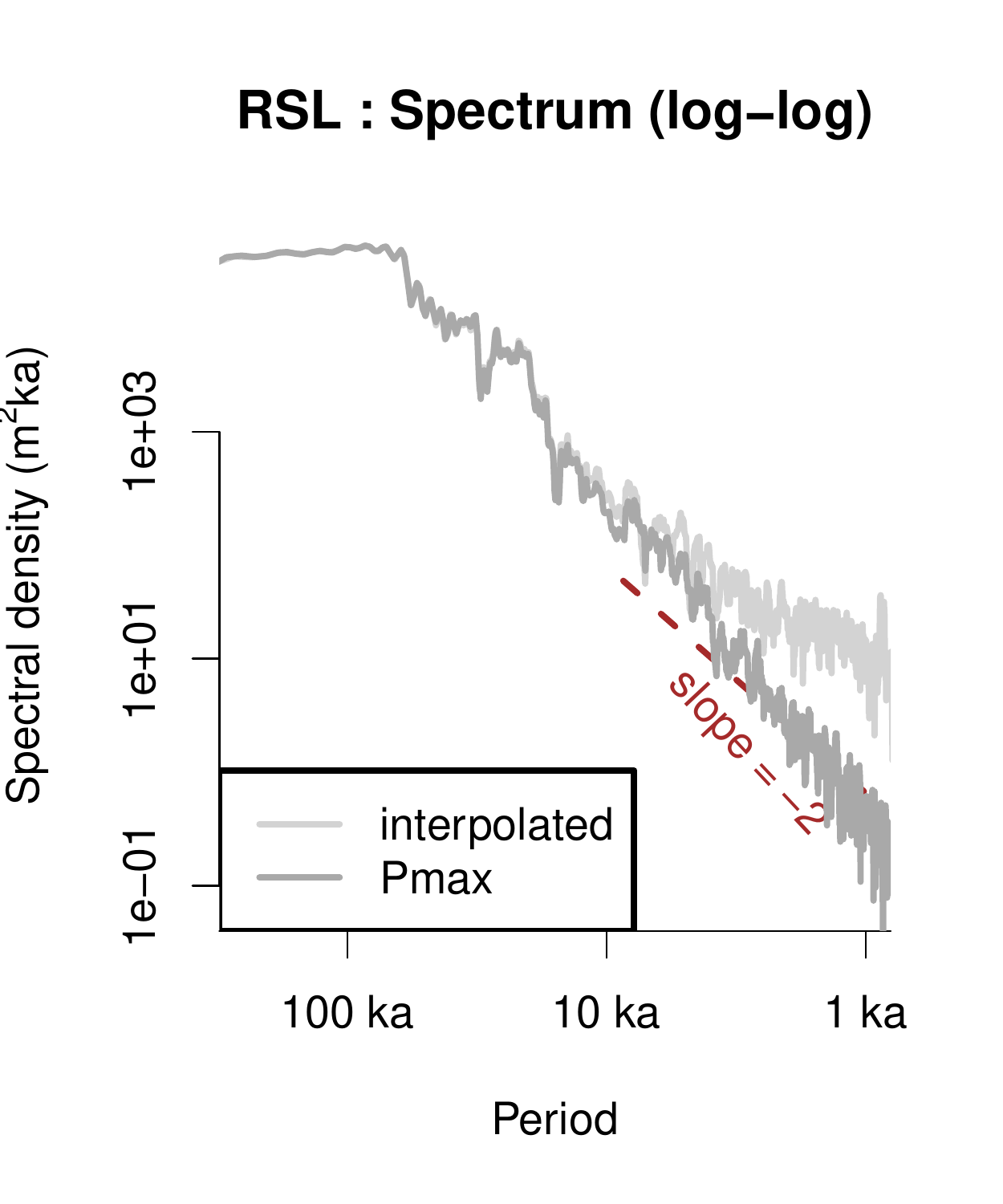}
\caption{
Relative Sea level (RSL) from Red Sea sediments \citep{Grant14aa} (a) Time series with raw data (black), linearly interpolated data with a 100yr time step (light gray), and maximum of probability (`Pmax') data at constant time step (dark gray). All data are supplied in (Grant et al., 2014). 
(b) Spectra of interpolated and Pmax data using a one ``taper"-tampering method, following \cite{percival98} (c) Adaptive (weighted) multi-taper spectra of interpolated and Pmax data, with 6 tapers (analysis carried out with the « pmtm » Matlab function \citep{MATLAB_R2013a} after average removal).  A line with slope $-2$ is added for visual reference. The multi-taper was developed to estimate the high-frequency tails of spectra. The multiple tapers reduce spectrum variance as well as the spectral leak, but they also flatten spectral peaks, hence the necessity, e.g., outlined by \cite{Ghil02aa}, to complement the analysis with other techniques focusing on variance modes. 
}
\label{fig:sea-level}
\end{figure}

\begin{figure}
\includegraphics[scale=0.6]{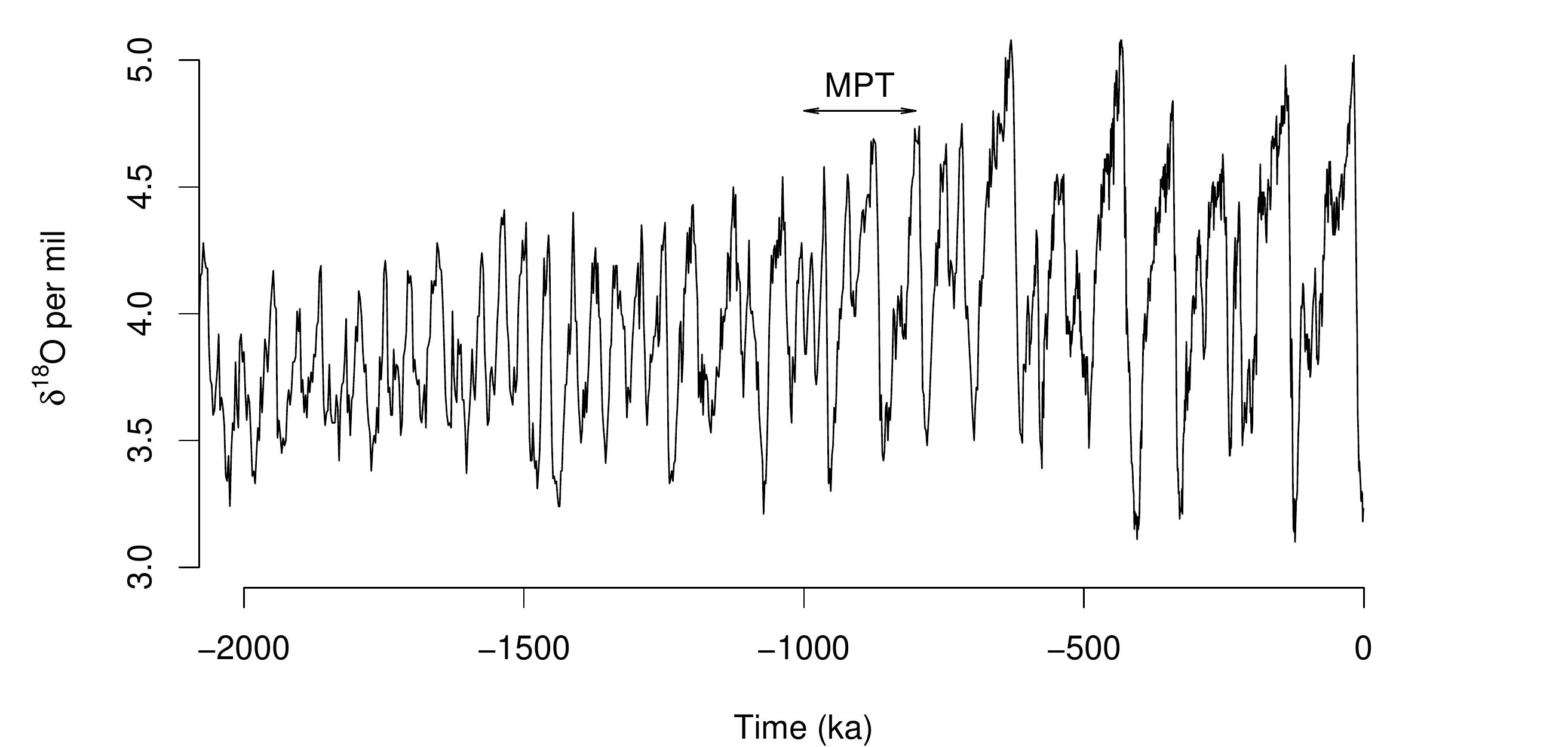}
\caption{
The  \citet{lisiecki05lr04} `stack' is a composite record of 57 series of isotopic ratio of oxygen isotopes measured in shells of benthic foraminifera. It is commonly used as a proxy for global climate conditions, and shows well the increase in amplitude and length of ice age cycles over the last two million years.
}
\label{fig:lr04}
\end{figure}

\begin{figure}
\begin{tabular}{ll}
\textsf{\quad (A) Saltzman (1990)} & 
\textsf{\quad \quad (B) Huybers and Curry (2006)} \\
\\
\raisebox{2em}
{
\includegraphics[scale=0.4]{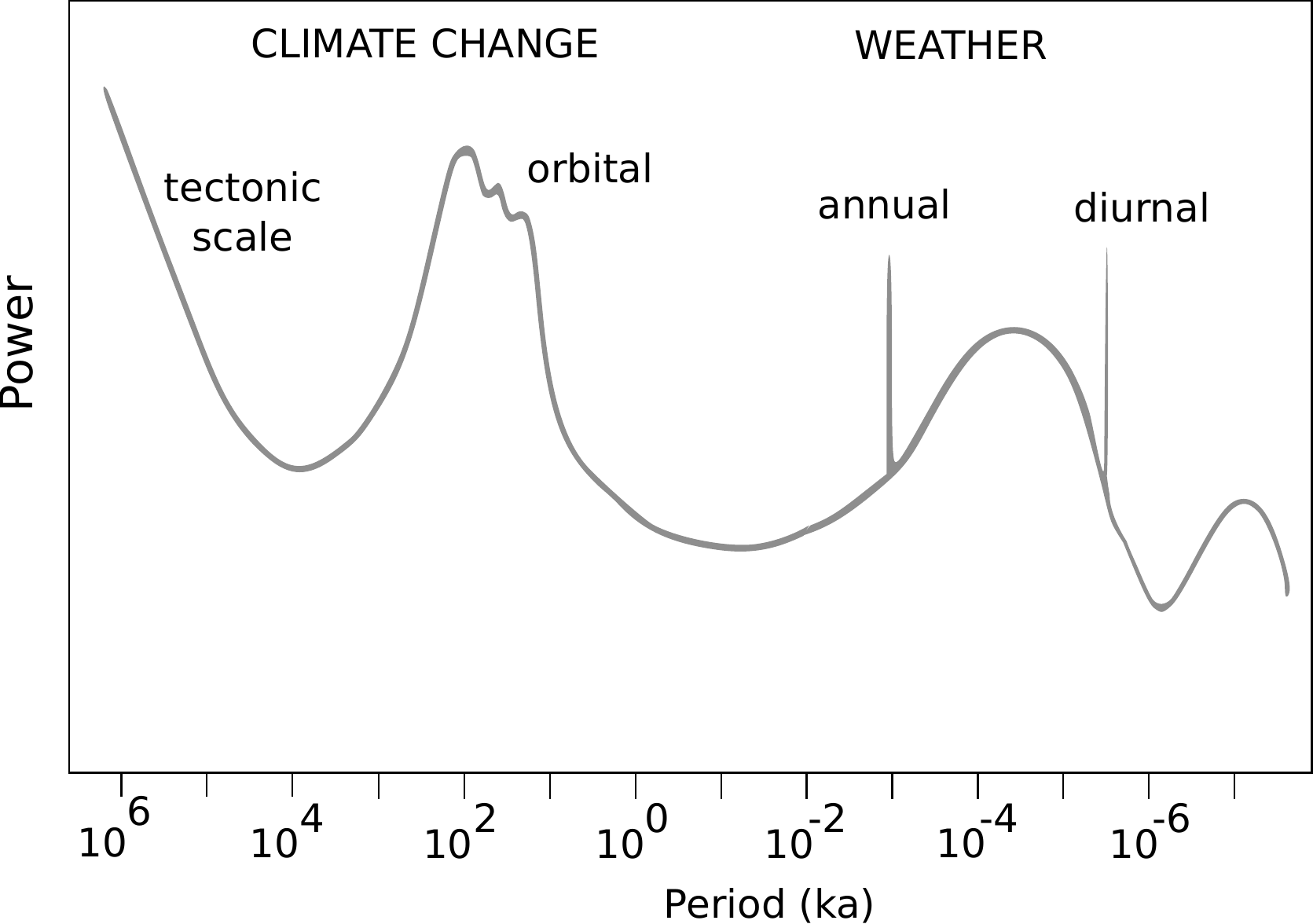}
}
&
\includegraphics[scale=0.4]{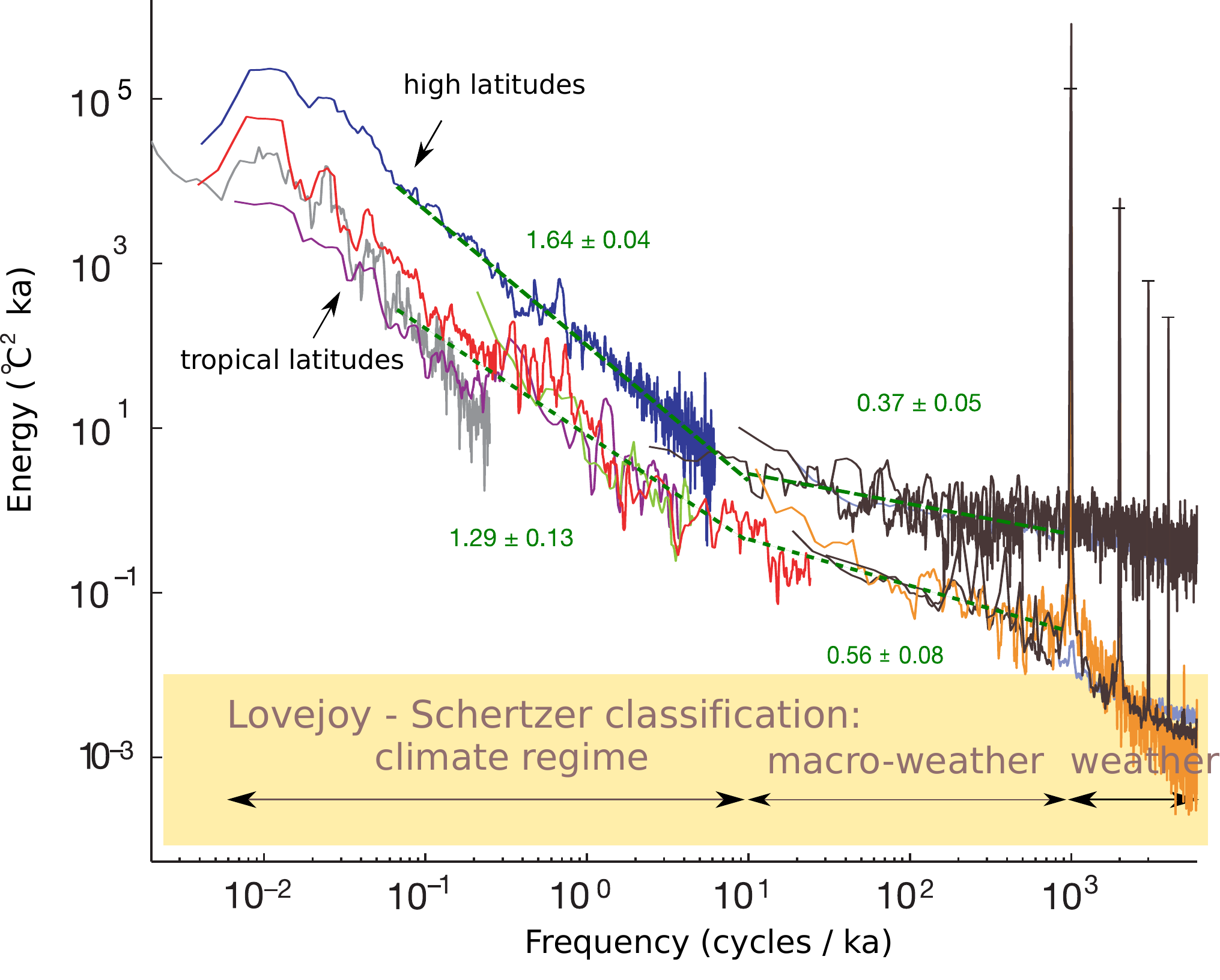}
\end{tabular}
\caption{Two previously published temperature power spectra.   \citet{Saltzman90aa} meant to be highly idealised, while \citet{Huybers06aa} is quantitative. Note the important structural difference: Saltzman distinguishes \textit{climate} and \textit{weather} regimes localised in the spectral domain and separated by a~gap, giving full justification to the time-scale separation needed for the \citeauthor{Hasselmann76}'s stochastic theory (\citeyear{Hasselmann76}).
The spectrum provided by \cite{Huybers06aa} is presented in log-log form.  These authors analysed different data types and estimated spectral slopes (in green). 
On this plot, the more-energetic spectral estimate is from high-latitude continental records and the less-energetic estimate from tropical sea surface temperatures. 
The different data types are marked with the color codes. Recent data come from instrumental records and re-analyses. Other data are from various natural archives (see the original reference for details). Note that the statistical estimation method of spectral slopes was not specifically adapted to unevenly spaced time series. 
The classification of regimes suggested by \citet{Lovejoy12aa} is added in yellow. Figures were adapted from the original publications.}
\label{fig:spectrum}
\end{figure}

The ice ages \index{ice ages} may be characterised as follows:

\begin{enumerate}
  \item This is a~global phenomenon.
The waxing and waning of Northern Hemisphere ice sheets is coupled to variations of sea-level of the order of one hundred meters (Figure \ref{fig:sea-level}A); it is also associated with large and significant variations in greenhouse gases concentrations (CO$_2$, CH$_4$ and N$_2$O), tropical atmosphere dynamics, land cover and ocean circulation \cite[e.g.][]{Ruddiman06ab}.
Different records provide complementary information on the different facets of the climate response, and considerable attention is being paid to the chronological sequence of this response \citep{Imbrie92aa,  shackleton00, Ruddiman06ab,  Lisiecki08ac}.
  \item Ice ages combine fast and slow fluctuations.
For example, the time elapsed between the latest two interglacial conditions  is of the order of 100\unit{ka}. 
During  the latest deglaciation, CO$_2$ concentration increased by 50\unit{ppm}, that is about 50 \% of its glacial-interglacial range, in only 3\unit{ka} \citep{Monnin01aa},  and by 10\unit{ppm} in 200 years  \citep{Marcott14aa}. 

  \item The frequencies and modulation patterns of the elements of the astronomical forcing (defined in section \ref{sect:astro}) are well recognised in climate records \citep{hays76, lisiecki07trends} (Figure \ref{fig:sea-level}B).  There is also a~statistically significant relationship between the timing of deglaciations, and the timing of astronomical configurations causing larger-than-average incoming solar radiation (insolation) in the Northern Hemisphere in summer \citep{Raymo97aa, Huybers11aa}, although not all positive insolation anomalies give rise to a  deglaciation.
  \item The duration and amplitude of ice ages has been varying through time.
The latest four cycles are characterised by a~marked saw-tooth shape pattern, with slow glaciation and relatively rapid deglaciation.
They cover the last 400\unit{ka} \citep{broecker70}. 
Before 900\unit{ka}~BP, climate fluctuations generally followed a~cycle of the order of 40\unit{ka} \citep{ruddiman86}. The lengthening of ice ages around 900\unit{ka}~BP is called the Mid-Pleistocene Transition \index{Mid-Pleistocene Transition} \citep{Clark06aa} (Figure \ref{fig:lr04}). 
The 100\unit{ka} periodicity dominates the spectrum  \index{power spectrum} of ice age fluctuations  since the Mid-Pleistocene Transition. These cycles display a pattern of modulation of amplitude and frequency that is compatible with a form a non-linear synchronisation \index{synchronisation} on eccentricity \citep{Rial04aa, Berger05aa, Rial13ab} (see again section \ref{sect:astro} for definitions of astronomical elements).
\item 
Ice ages represent a mode of climate variability connected to variability  at shorter  and longer time scales. While seemingly obvious, this statement
 represents a shift in our understanding of the climate spectrum. In the past  
\citep[e.g.][]{saltzman90sm}, it was suggested to clearly distinguish `weather' and `climate phenomena' as distinct modes of variability, decoupled and separated by a spectral gap (Figure \ref{fig:spectrum}). The postulated spectral gap would have acted as an efficient information barrier between climate and weather processes. 
Closer inspection of the data power spectra shows however that the postulated gap does not exist \citep{Pelletier97aa, Huybers06aa, Lovejoy12aa}. 
The power spectrum of temperature is estimated to have approximately the typical shape of a~Lorentzian spectrum, of which the power increases from the millennial scales, up to  40\unit{ka}, where it flattens (\citet{Pelletier97aa}, Figure shows  \ref{fig:sea-level}C a sea-level power spectrum and Figure \ref{fig:spectrum}B a temperature one).
On the other hand, the spectrum flattens significantly at time scales of 1~ka and below, with only a moderate slope up to the annual time scale. This latter range of variability has been named the `macro-weather' regime by \cite{Lovejoy12aa}. \index{macro-weather}
\end{enumerate}

\subsection{The astronomical forcing \label{sect:astro}}
\index{astronomical forcing} \index{Milankovitch}
The astronomical forcing (also referred to as orbital or Milankovitch forcing) is the action of the slow changes in Earth's orbit shape, position of perihelion, and obliquity on the seasonal course of incoming solar radiation (insolation) hitting the top of the atmosphere at any point on Earth \index{insolation}.
To an excellent approximation, the great semi-axis of the Earth's orbit is constant \citep{Lagrange1781,Laskar04}. Hence, the astronomical forcing may be determined from variations in Earth's orbital eccentricity $e$, the true solar longitude of the direction of Earth's perihelion ($\varpi$), and obliquity ($\varepsilon$).

The total amount of energy received all over the globe in one year is proportional to $1-e^2/2 + \mathcal{O}(e^3)$.
Accounting for the fact that the eccentricity varies between $0$ and $0.05$, this represents fluctuations of no more than $0.1\%$ of the absolute value of insolation. \index{insolation}
On the other hand, changes in obliquity affect the seasonal contrast of insolation as well as the latitudinal distribution of annual mean insolation. Changes in the longitude of perihelion, modulated by eccentricity, produce monthly anomalies compared to the averaged seasonal cycle because the amount of insolation is inversely proportional to the squared distance to the Sun. The annual mean anomalies associated with obliquity are then of the order of 5 W/m$^2$, and the monthly-mean variations associated of precession can be of the order of 50\unit{W/m^2}. 
It is therefore generally recognised that the astronomical forcing affects Earth's climate because it changes the seasonal and spatial distributions of insolation. 

In low-order or conceptual models (see section \ref{sect:conc}), the astronomical forcing  is often summarised by one or two well-chosen forcing functions. 
For example, \citet{milankovitch41} used  insolation over the \textit{caloric summer}. 
This is the amount of insolation integrated over the half of the year experiencing the largest amount of insolation.
The quantity may be used as a~predictor of the amount of snow susceptible of surviving the summer season, and therefore be used  as a~forcing term in ice sheet-climate models.
Alternative forcing functions include insolation at 65$^\circ$ N at the summer solstice or in July  \citep{imbrie80,  saltzman90sm}.
These quantities are particular cases of a~wide class of insolation forcing functions that are approximately a~linear combination of $e\sin\varpi$, $e\cos\varpi$ and $\varepsilon$ \citep{Loutre93aa,  Crucifix11aa}.  The
power spectrum of these quantities is useful to know, because the astronomical forcing is believed to excite the dynamics of the climate system through this spectrum. 

\begin{figure}
\input{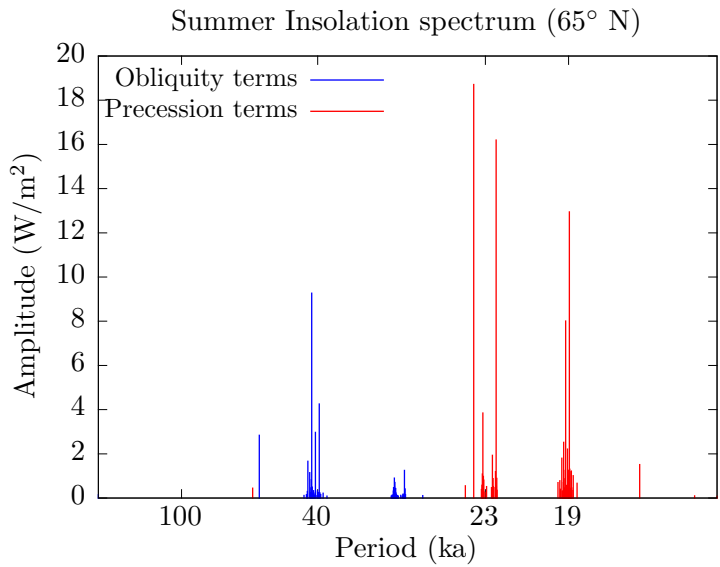}
\caption{
\label{fig:insol}
Analytical spectrum of summer solstice incoming solar radiation at 65$^\circ$ N, assuming an approximation of this insolation as a~linear combination of $e\sin\varpi$,  $e\cos\varpi$,and $\varepsilon$ ($e$: eccentricity, $\varpi$, true solar longitude of perihelion, and $\varepsilon$: obliquity), and the analytical development of \citep{Berger91aa}.
}
\end{figure}

Nowadays, the astronomical forcing is known from numerical solutions of an~eleven-body problem  (Sun + 9 planets + Moon ) \citep{Laskar04, Laskar11aa}.
However, the earlier solutions of \citep{berger78} and  \citep{Berger91aa}
are still used because their spectral decomposition is given explicitly (Figure \ref{fig:insol}). 
The dominant periods are \citep{Berger91aa}:
\begin{enumerate}
  \item for precession ($e\sin\varpi$): 23708,  22394,   18964,  19116,  23149, \ldots years
  \item for obliquity ($\varepsilon$): 41090,  39719,  40392, 53864, 41811 \ldots years
\end{enumerate}

The spectral decomposition of $e$ may also be deduced from the expression  $e = [(e\sin\varpi )^2 + (e\cos\varpi )^2 ]^{1/2}$.
The dominant periods are then  404177,  94781,   123817,  130615, \ldots  years.

\section{Climate models used at the palaeoclimate scales}
We briefly review in this section a standard classification of climate models used in palaeoclimate research, into `conceptual', `earth models of intermediate complexity' and `global climate models'. \index{conceptual model}
\subsection{Conceptual climate models \label{sect:conc}}
\cite{claussen99} characterised conceptual climate models by the fact  that the number of free parameters has the same order of magnitude as the number of state variables. Consequently these models require little computing resources and they are typically integrated on personal computers. 
Conceptual models represent a very broad category.  One may distinguish  discrete and continuous dynamical systems, deterministic and stochastic systems, and the literature provides examples of the different combinations. 

Some of the early dynamical system models of ice ages were derived from ice-sheet flow equations  reduced to a~small number of ordinary differential equations \citep{Weertman76, Ghil81, letreut83}.
One of the objectives of these models was to explain the emergence of 100-ka climatic cycles as a~plausible consequence of ice sheet response to astronomically-controlled variations of the snow line.
Hence, the parameters used in these models were well identified physically, and this restricted the plausible range of these parameters.
These early models then evolved towards greater realism and complexity. Advances in computing power allowed to  resolve the partial differential equations on a spatial grid \citep{Oerlemans80, Birchfield81aa, Hyde85aa}. These models constitute the basis of modern state-of-the-art ice sheet models.

Since these early works, our knowledge of past CO$_2$ concentrations \citep{Delmas80aa, genthon87} has made it clear that ice age  modelling requires also to consider carbon-cycle dynamics.  Several dynamical system models since the eighties include prognostic variables for ocean and carbon pools, in order to account for the possibility of greenhouse gas exchanges between the oceans and the atmosphere \cite[e.g.][]{saltzman88,paillard04eps}.

As a~general rule, the description of biogeochemical mechanisms is partly speculative.
For example,  several models of the same authors \citep{saltzman88, saltzman90sm, Saltzman91sm} have slightly different equations for the carbon cycle dynamics. In  fact, conceptual models have been used to support various and partially conflicting interpretations of ice ages, emphasising the roles of ocean vertical mixing and sea-ice \citep{Gildor01aa}, 
alkalinity balance in the ocean \citep{Omta15aa}, or more generically feedbacks between temperature and CO$_2$ \citep{Hogg08aa}. 
The search for low-order models of ice ages is thus partly heuristic. It is risky to decide that one interpretation is better than another on mere inspection of the simulated sea-level  \citep{tziperman06pacing}.
With this caveat in mind, several investigators found it interesting to develop general theories that do not directly depend on a~specific physical mechanism, but which have implications on the predictability \index{predictability} or stability of ice ages. A couple of examples follow. 

\begin{figure}
\includegraphics[scale=1.0]{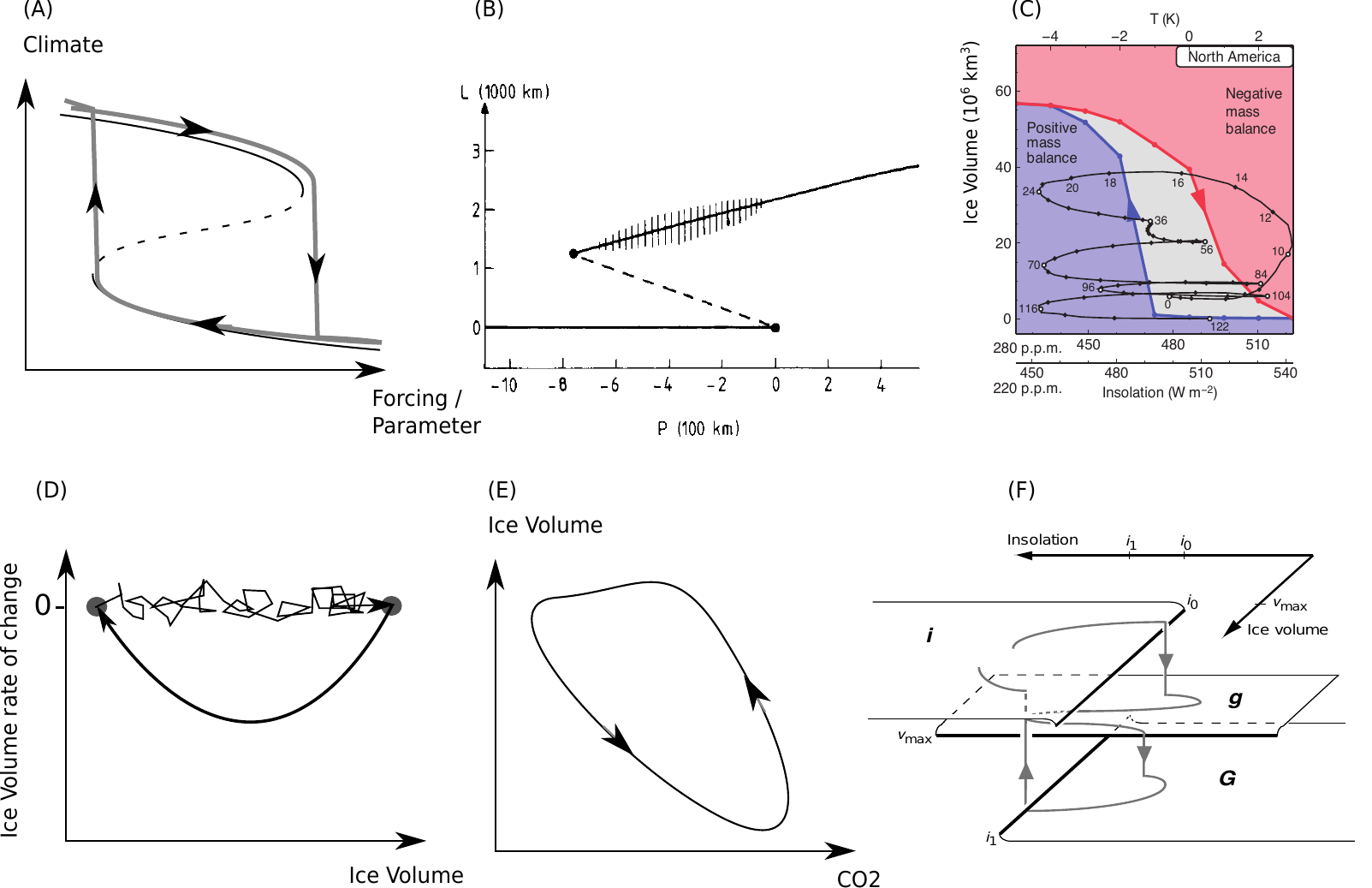}
\caption{
\label{fig:models}
(A): Generic representation of a~system with two stable states (separated by an unstable state, dashed) and two fold bifurcations. In a~more complex model, the two state branches can be estimated by means of a~hysteresis experiment (arrows, see also Section \ref{sect:emic}). (B): Experiments with a~simple ice sheet model \citep{Oerlemans81aa} broadly support this scenario. Hashes on (B) indicate an 'almost intransitive state', with sluggish dynamics; $P$ refers to the position of the snow line (free parameter) and $L$ the extent of the Northern Hemisphere Ice sheet. (C): same but with a~state-of-the-art model ice sheet atmosphere \citep{Abe-Ouchi13aa}. Thick colored full lines are the estimated system steady-states deduced from a~hysteresis experiment, and the thin black curve with numbers represent an actual simulation over the last ice age, numbers denoting time, in thousand of years before present. Figures (D, E, F) are possible mechanisms for transitions between glacial and interglacial states. (D): Stochastic accumulation, with a~flush mechanism to restore interglacial conditions;  (E): deterministic limit cycle, caused by interactions between different system components, (F): transitions forced by changes in insolation, in this case with the postulate of an intermediate state and asymmetric transition rules \citep{Paillard98}. Figures (C) and (F) reproduced by permission of Nature}
\end{figure}

\begin{enumerate}
\item A~possible starting point is to think in terms of multiple equilibria (Figure \ref{fig:models}A). 
If we assume that  feedbacks in the climate system combine to yield a stabilising response in response to a small external perturbation, at least at a certain time scale, then we come to the conclusion that the climate system is stable at that time scale. We may however speculate about some non-linear effects and imagine that a large perturbation will induce runaway effects. In this case, the climate system will drift away from its original state, towards another stable state. The co-existence of two stable equilibria appears in the energy balance model of climate proposed by \cite{Budyko69} and \cite{Sellers69aa}, as a consequence of a non-linearity introduced with the albedo feedback (Ditlevsen, in this volume). Nowadays, we know that the Budyko-Sellers theory has some relevance to study ancient glaciations such as those in the Ordovician but  it is not immediately applicable to ice ages. However, if we consider a thought experiment in which astronomical forcing \textit{and} CO$_2$ concentration would be constant, then, for certain values of such parameters, simulations with ice-sheet-atmosphere models suggest to us that two stable states may effectively coexist
(\cite{Oerlemans81aa}, \cite{Calov05aa}, and \cite{Abe-Ouchi13aa}, Figures \ref{fig:models}B and C). 
Having multiple equilibria is however not enough to explain ice ages:
A mechanism is needed to jump from one state to the next.
Historically, it was proposed to involve a~mechanism of stochastic excitation \index{stochastic excitation} causing transitions between the two states. 
More specifically, two stochastic mechanisms leading to ice ages were proposed:
stochastic resonance \citep{BENZI82aa, NICOLIS82ab,  Matteucci89aa} and coherence resonance \citep{Pelletier03aa}. 

Stochastic processes can be justified as a representation of fast internal variability such as chaotic variability associated with atmosphere and ocean dynamics. 
In \textit{stochastic resonance}, \index{stochastic resonance} state transitions are induced by these additive fluctuations but their timing is controlled by a~weak external forcing. 
This  theory requires forcing at 100\unit{ka}, but this is not a strong frequency in the astronomical forcing. Furthermore, it produces ice age curves with rapid and broadly symmetric transitions between glacial and interglacial states. This conflicts with observations. 
\textit{Coherence resonance} addresses these shortcomings \index{coherence resonance}.  In this case, the oscillation mechanism relies on a~delay-feedback which, in \cite{Pelletier03aa}'s model, depends on the rate of climate change.
This feedback accelerates the deglaciation process once started. 
\item Stochastic dynamics can generate ice age cycles even without the requirement of stable states.
\cite{Wunsch03spectral} shows this with a very simple model. 
The drift from interglacial towards glacial conditions is modelled as a random walk, which terminates  by an abrupt flush towards the interglacial state when glaciation exceeds a threshold (Figure \ref{fig:models}D). 
The physical nature of the rapid deglaciation is itself uncertain. 
\cite{Pelletier03aa} theorised on the ice sheet mechanical collapse, a possibility supported by ice-sheet experts \citep{Pollard83aa, Abe-Ouchi13aa}. A flush may also be induced by the dynamics of the  southern ocean / carbon cycle dynamics. This second possibility is favoured by \citet{paillard04eps} and \citet{Paillard15aa}, though these authors favour a deterministic framework. 
\item
It is also possible to obtain alternation of glacial and interglacial climates as a result of non-linear interactions between different components of the climate system, without the requirement of an external forcing. In dynamical systems language, this amounts to view ice ages as a deterministic limit cycle (Figure \ref{fig:models}E). 
Saltzman et al. developed this theory in a series of articles, including \cite{Saltzman81aa,saltzman88, saltzman90sm,  Saltzman91sm}.
One of the advantages of the limit cycle concept is that it provides a~natural way to explain the saw-tooth shape of ice ages \citep{Gildor01aa, ashkenazy06phase, paillard04eps, Ashwin15aa}. It also provides a promising starting point to think about the  dynamics of the Mid-Pleistocene Transition because this transition may then be interpreted as a bifurcation \citep{Saltzman90aa,Crucifix12aa,Ashwin15aa,Mitsui15ad}. 
The effect of astronomical forcing may be accounted in limit cycle models \index{limit cycle} as a small additive forcing, which controls the timing of glaciation and deglaciation events, through a~phenomenon of synchronisation \citep{tziperman06pacing}.  \citet{De-Saedeleer13aa}, \cite{Crucifix13aj} and \citet{Mitsui13aa} \label{page:struct_inst} studied such synchronisation \index{synchronisation} mechanisms in detail
and they found that, with the astronomical forcing, the attractors of oscillator-type models of ice ages may become strange and non-chaotic \index{strange non-chaotic attractor}. 
In simple terms, the modelled trajectories are synchronised to the astronomical forcing, and their greatest Lyapunov exponent (conditioned on the forcing) is slightly negative. In this rather unusual regime,  the dynamics are not chaotic, and different initial conditions will typically converge to the same trajectory.
However, the dynamics associated with strange non-chaotic attractors are such that the exact ice age trajectory is fragile with respect to small changes in the parameters, or to the addition of a~weak stochastic process (Figure \ref{fig:sm91}).
The nature of the bifurcations from quasiperiodic to strange nonchaotic attractors \index{strange non-chaotic attractor} are studied in more details in \citet{Mitsui15ad} on the basis of a~phase oscillator.

\item Finally, ice ages may be viewed as a forced oscillation \textit{driven} by the astronomical forcing. 
A~particularly clear  example is provided by \cite{Paillard98}, who defines three states (glacial, semi-glacial, interglacial) and transition rules involving astronomical forcing and ice volume thresholds (Figure \ref{fig:models}F).
The idea was reformulated in terms of continuous dynamics in \citet{Ditlevsen09aa}, and then associated to a~mechanism of relaxation oscillation in \citet{Ashwin15aa}.  A~number of publications  focus specifically on the identification and analysis of threshold functions \citep{Parrenin03aa,Parrenin12ab, Feng15aa} 
\end{enumerate}

The above examples do not exhaust the possible interpretations of ice age dynamics in terms of dynamical systems concepts.
\citet{letreut83}, and \citet{Daruka15aa} emphasised the notion of non-linear resonance, \citet{Roberts15aa} developed an ice age theory featuring  mixed-mode oscillations associated with Canard trajectories, and   \citet{rial99,  Rial04aa, Rial13ab} interpreted ice ages in terms of frequency modulation theory, emphasising the role of eccentricity.
These models may not all be equally relevant, but each of them unveils possible counter-intuitive effects of the  astronomical forcing on non-linear climate dynamics.
They are paradigmatic, in the sense that they provide  research questions \textit{and} conceptual frameworks, which can be referred to when experimenting with models of higher complexity. 

\subsection{Earth Models of Intermediate Complexity \label{sect:emic}}
Parallel to efforts with conceptual models, we need to explain emergent dynamics of ice ages from equations describing physical processes such as ice and water flows, vegetation dynamics, sea~ice, etc. 
Detailed representations of all these processes would require considerable algorithmic complexity. 
Hence, in practice, any numerical model of the Earth System is characterised by a  balance between what is \textit{simulated} as an emergent property, and what is parameterised.
This balance determines the resulting level of model complexity.

In some of these models  \citep[e.g.][]{gallee91, petoukhov00}, the atmospheric mesh is too coarse to capture baroclinic activity. 

These models therefore rely on parameterisations 
to account for the aggregated effects of \textit{weather} on the average flows of heat, momentum and water. 
The parameterisations are semi-empirical diagnostic and deterministic models written to 
satisfy theoretical considerations and accommodate observations. Pioneering references for heat and moisture transport parameterisations include \citet{Saltzman71aa}, \citet{Sela71aa}, and more recently \citet{Barry02aa}. 
In some cases  \citep[again][]{petoukhov00} ocean dynamics are zonally averaged in each ocean basin, and  the relationship between the zonal and meridional pressure gradients in the ocean is also determined by a parameterisation. The 
gain in computing time induced by these simplifications permits integrations over several tens of thousands of years, and to account for the dynamics of slower climate system components such as ice sheet, carbon and sediment dynamics. These dynamics are needed to generate the ``climate regime" indicated on the Figure \ref{fig:spectrum}. 
Models of this kind are called `Earth System Models of Intermediate Complexity' (EMICs) \index{EMIC} \citep{claussen99}. 
The category of EMICs is pretty broad and some EMICs resolve ocean and atmosphere more explicitly. For example  LOVECLIM \citep{Goosse10aa} simulates some of the eddy atmospheric transport and include 3-dimensional ocean dynamics, but it also includes  strongly simplifying assumptions about atmospheric convective activity and cloud dynamics. 

Even though these models remain idealised in many aspects, the jump in algorithmic complexity from conceptual models to EMICs is considerable.
\citet{claussen99} observed that the state vector of an EMIC has orders of magnitude more dimensions than the number of parameters. 
One of the consequences of the high dimensionality of the state vector, is that the bifurcation structure may be difficult to comprehend. 
It would however be reassuring to verify that dynamical systems concepts developed under the theoretical framework of conceptual models remain relevant in high-dimensional systems. 
For example, 
the climate model of \cite{Sellers69aa} has a~bifurcation between cold and warm states \cite[see also][]{Ghil76aa}. 
Is this characteristic actually useful to describe the real world? 
Hysteresis diagrams (Figure \ref{fig:models}C) are one attempt to answer the question. 
The simple principle behind the construction of such diagrams is the following: a~control parameter is changed slowly forward and backward in the model in order to uncover possible irreversible transitions. 
Such transitions may then be interpreted as the manifestation of a bifurcation.
This experiment was carried out for different ocean \citep{Rahmstorf05aa}, vegetation \citep{Brovkin98aa}, and climate ice-sheet models \citep{Oerlemans81aa, Calov05aa,Abe-Ouchi13aa} (this last example was based on  a more complex GCM coupled to an ice-sheet model). \index{GCM} \index{hysteresis diagram}
The hysteresis diagram was once termed a~form of `poor's man continuation algorithm' \citep{Dijkstra05ab}. 
It may indeed appear as a~rudimentary process to characterise dynamics that may in fact by quite complex, and
there are other possible ways to characterise EMIC dynamics that have remained largely unexplored. 
For example, methods exist to reconstruct bifurcation diagrams for high-dimensional systems \citep{Dijkstra05aa}. 
Techniques are also being proposed to probe stable and unstable manifolds more systematically, for example by implementing a~control loop to stabilise the unstable manifold \citep{Sieber14aa}. 

\subsection{Global Climate Models \label{sect:gcm}} \index{GCM}
The phrase Global Climate Model is a~rewording of the original acronym of General Circulation Model.
GCMs were originally based on numerical weather forecast models, but 
 the typical modern GCM includes ocean, sea-ice, vegetation, carbon-cycle dynamics, ocean and atmosphere chemistry, aerosols and other components.  
\textit{Sensu stricto} an EMIC is also a global model of the climate system, but to comply with a widespread usage we reserve here the word GCM to designate a model 
used to generate \textit{weather}  and \textit{macro-weather} statistics, which  EMICs do not do, or at least not so well. 
GCMs also include parameterisations, but they must account for smaller scale processes than parameterisations in EMICs. This includes cloud cover dynamics, convective plumes, gravity waves and, as in EMICs, these  parameterisations are generally expressed as deterministic semi-empirical models \citep{Palmer05aa, Stensrud09aa}. 
Because the research agenda in climate science is strongly  influenced by programmes focusing on next century's climate, state-of-the-art GCMs tend to have the maximum resolution affordable on state-of-the-art civil computers to carry experiments of ca. 100 to 500 years. In palaeoclimate research, state-of-the art GCMs are mainly used to produce short experiments with fixed boundary conditions and parameters: such experiments are called snapshot or time-slice experiments. GCMs of the previous generation, which can be run for a couple of thousand of years (at the price of several months of computing time) are used for long equilibrium experiments \citep{braconnot07pmip1} or transient experiments during which boundary conditions are allowed to change to simulate specific events such as the last deglaciation \citep{He13aa}, or even the last glacial-interglacial cycle if spacial resolution is reduced \citep{Smith12ab}.

\section{Why we need to combine modelling frameworks}
The key challenge raised by palaeoclimate observations may be summarised as the task of accommodating existing observations of different nature,  associated with processes at different time scales, in a consistent framework. This task requires considerable insight about physical and biogeochemical processes and, as we  see throughout this chapter, it has proved to be highly non-trivial. Opportunities also exist for predicting palaeoclimates signals even before they will be observed. 
For example, CO$_2$ atmospheric concentrations are currently known back to 800\unit{ka} BP \citep{Luethi08aa}, but it is suspected that the Antarctic ice sheet conceals ice that is at least 1.0 Ma old \citep{Parrenin15aa}.  A~prediction of the CO$_2$ concentration before 800\unit{ka} would  be subject to testing when this old ice is recovered and analysed. 
The spatio-temporal properties of centennial and millennial variability \index{centennial variability} are also poorly characterised, especially in tropical areas and before the last glacial-interglacial cycle, but such data are becoming available. For example,  a~current drilling project is hoped to deliver 200\unit{ka} of high-resolution climate data~in central Africa \citep{Verschuren13aa}. Such records will thus provide further opportunities to test our understanding of climate dynamics. 

Identifying interactions and predicting observations require models. 
As we have just seen, some models are more complex than others, but  \textit{any} model includes a fair number of decisions that are not immediately determined by physics, biology or chemistry, starting with the decision to account or not for a system or a specific process. Parameterisations \index{parameterisation} are typical places where scientists  inject judgement about how a given process should be represented, and these judgements are subject to revision if the model dynamics do not accommodate observations well enough. 

The heuristic and uncertain nature of parameterisations cannot be overstated. Meteorologists understand this \citep{Palmer05aa}, but they tend  to focus on uncertainties about sub-grid scale processes, such as convection and cloud formation. 
As we are interested here in palaeoclimates, we should also bear in mind that there are large uncertainties about the biological and physical processes at the palaeoclimate, global scale
\citep[][chap. 15]{Saltzman90aa, saltzman02book}:
How  biogeochemical cycles respond to ice ages is not so well known. Ice-sheet physics remain also partly elusive. 
These interactions need to be characterised by inspection of the data, in part through a process of trial and error. As a general rule, it is good practice to test hypotheses with parsimonious models that include few free parameters. This explains the interest of some palaeoclimate scientists for conceptual models. One advantage of low-order dynamical systems models is also that they provide a means to express our understanding of the slow dynamics of the climate system by reference to  concepts pertaining to dynamical systems theory.  

On the other hand, \cite{Lorenz70aa} envisioned  that with enough computing power `` in the 21st century [\ldots] we may construct a~super model [\ldots]   When we integrate the equations, if they are correct,  we shall necessarily obtain changes in climate, including the great ice ages. "
There is thus a tension between, on the one hand, the modelling needs created by our lack of knowledge about large scale palaeoclimate processes and, on the other hand, this vision of ice ages as a seamless extension of weather phenomena towards the low frequencies. Lorenz certainly appreciated this tension because he was careful enough to write the caveat ``if the equations are correct" in his statement. It is thus necessary to combine different modelling frameworks. We must use models of geophysical flows---including GCMs---to explain how variability modes emerge from physics, and we need lower-order models to test hypotheses on dynamical processes.
The remainder of this article is dedicated to strategies for synthesising information yielded by the different modelling approaches. 

\section{A research agenda}

\subsection{Modelling interactions between ice ages and the millennial variability \label{sect:seamless}}
We have seen above that a number of conceptual models explain the emergence of ice ages as a consequence of interactions between different components of the climate system, forced or controlled by the astronomical forcing. However, the question of interactions between ice ages and millennial variability is one that has not been addressed successfully. 

\cite{Wunsch03spectral} observed  that the shape of the ice age spectrum is broadly similar to a Lorentzian \index{Lorentzian spectrum}: ice age dynamics can thus be interpreted as the result of an auto-regressive (hence, stochastic) process (see \cite{PISIAS81ab} for earlier considerations along the same line). This view downplays the role of the astronomical forcing as a driver of ice ages.  \cite{tziperman06pacing} then qualified this statement. They  noted that the astronomical forcing can still act as a pacemaker through  a phenomenon  of synchronisation \index{synchronisation} (which they called ``phase-locking"). These authors suggested that the synchronisation may be effective even with fairly high levels of noise, compatible with the Lorentzian aspect of the spectrum, but we believe that this is not a closed issue. The robustness of the sequence of ice ages will effectively depend on the model (\cite{Crucifix13aj}), on the level of noise, and, presumably, the nature of the noise process.

\begin{figure}
\includegraphics[scale=0.5]{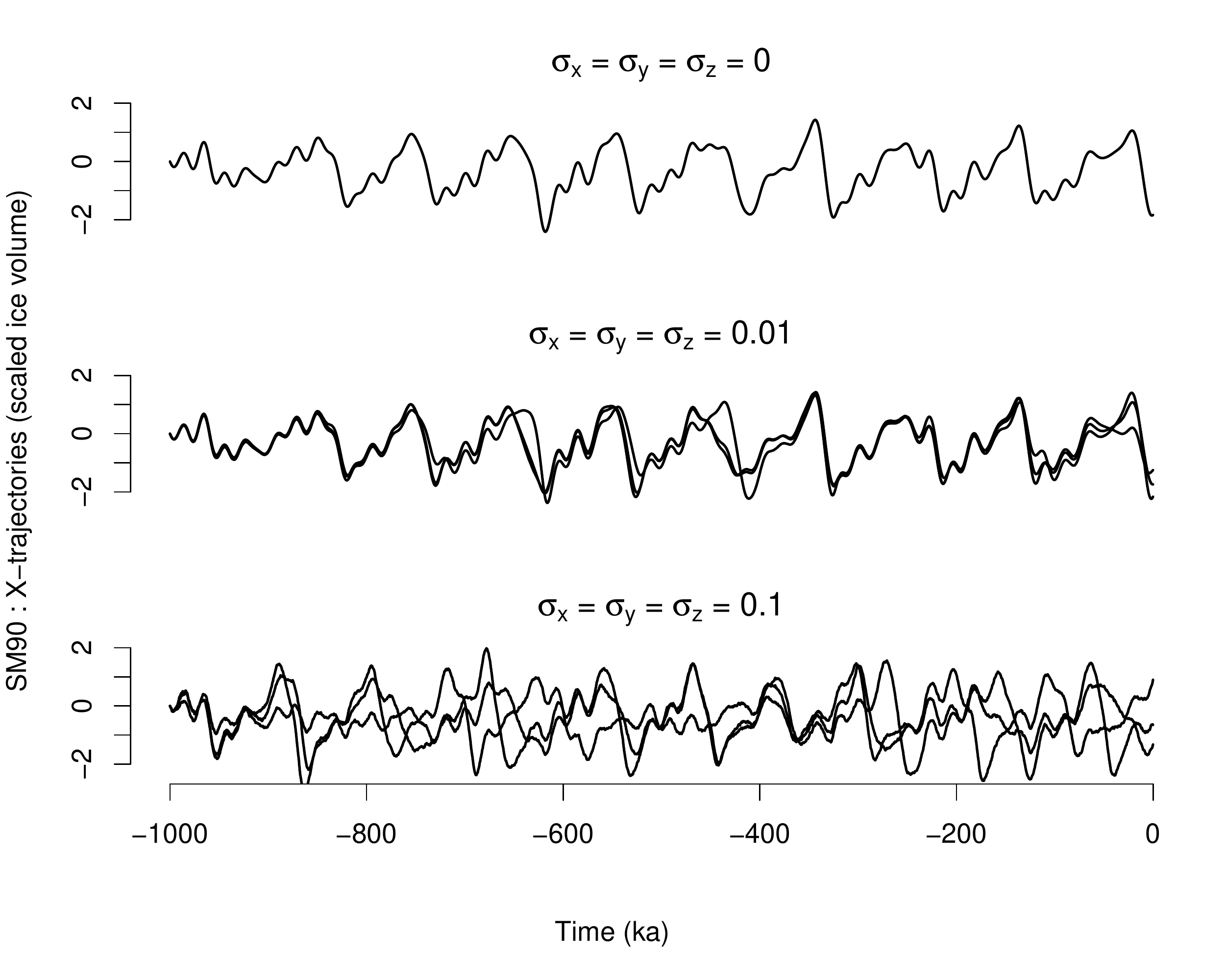}
\includegraphics[scale=0.5]{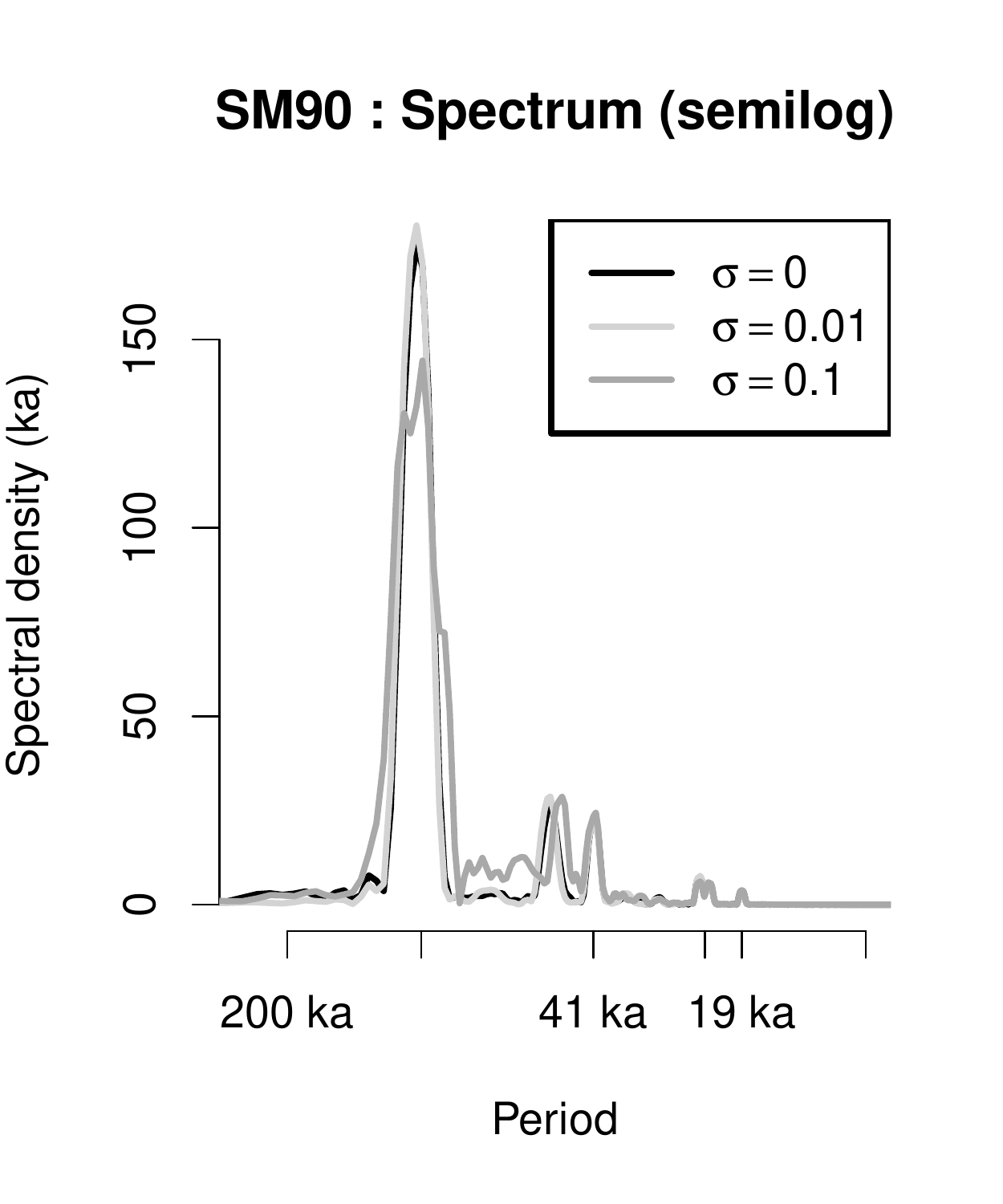}
\includegraphics[scale=0.5]{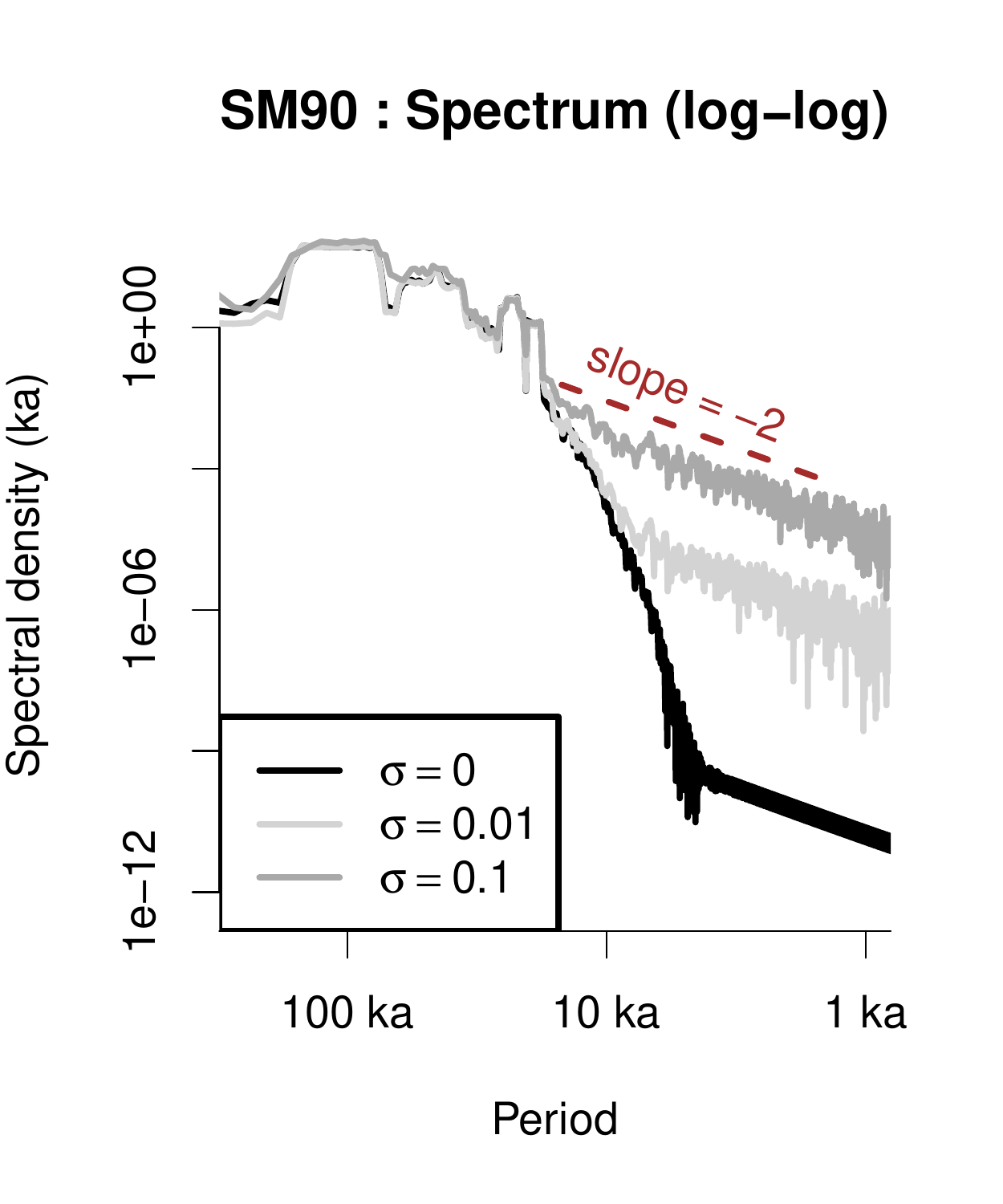}
\caption{The SM90 \citep{saltzman90sm} model is presented in the form of three dimensional equations $\dot X = - X - Y - vZ - uF(t) + \sigma_x \dot \omega_x$, $\dot Y = -pZ + rY - sZ^2 - wYZ - Z^2Y + \sigma_y \dot \omega_y$, $\dot Z = -q(X+Z) + \sigma_z\dot \omega_z$, where $\dot X$ is the time-derivative of $X$, $\omega_{x,y,z}$ are three independent Wiener processes, $F(t)$ is the normalised summer-solstice insolation at 65$^\circ$ North, and $u,v,w, p,q,r,s$ are parameters as given in \cite{saltzman90sm}. We provide (a) sample time trajectories (with different noise realisations) 
(b) Classical multi-taper spectra of interpolated and Pmax data, with 1 taper (c) Adaptive (weighted) multi-taper spectra of interpolated and Pmax data, with 6 tapers. Multi-taper analysis was performed with the « pmtm » Matlab function \citep{MATLAB_R2013a} after average removal.  
A dashed line with slope $-2$ is added for visual reference.
\label{fig:sm91}
}
\end{figure}

We further illustrate this point with  one of the models developed by the Saltzman school \citep[][SM90]{saltzman90sm}. 
It features three differential equations, representing ice volume, carbon dioxide concentration, and deep-sea~temperature (see caption of Figure \ref{fig:sm91}). 
In this model, the astronomical forcing is included as an additive term in the ice volume equation.
Weather variability is accounted for with three Gaussian white noise processes of constant amplitude, added to the different equations. 
We reproduced this model and  tried different noise amplitudes, including zero (Figure \ref{fig:sm91}). 
As the white noise term is integrated over time, it leaves a~specific signature on the spectrum in log-log scale, recognisable as a tail with slope of $-2$ connecting millennial variability with astronomical variability. 
The largest noise amplitude is needed to connect continuously the slope-$2$ tail with the forced and internal variabilities as the data suggests. 
However, in this specific case, the trajectories generated by the model become more random: Different noise realisations produce different sequences of ice ages.

In order to further investigate the issue of the effects of centennial and millennial variability on ice ages, it seems  necessary to consider the nature of the millennial variability in more detail. 
In particular, methods for estimating spectral power such as the multi-taper method  used in \cite{Wunsch03spectral} are suitable for estimating spectral tails by reference to stochastic models. Other time series techniques are however better suited to characterise modes of variability. \cite{Ghil02aa} reviews a few of them, including the continuous wavelet transform and the singular spectrum analysis. Empirical mode decomposition is another one \citep{Huang98aa}. Such methods started to be applied on palaeoclimate records in the 1990's \citep{YIOU94aa}, and they revealed the existence of millennial-scale dynamical modes. That observation had been predicted by a simplified atmosphere-cryosphere model \citep{Ghil81}. The authors of this model also inferred that millennial modes are necessary to generate 100-ka variability through a phenomenon of non-linear resonance \citep{letreut83}. That model, among others, correctly reproduces a slope-2 spectral structure, but it does not reproduce the sequence of ice ages. Once more, we face the contrast between two competing views on ice ages modelling, one emphasising the spectrum, the other its phase structure. These two views  need to be unified. 

We know today that climate millennial variability is in part characterised by typical events that leave a distinctive signature on the climate records. Dansgaard-Oeschger events \index{Dansgaard-Oeschger events} \citep{dansgaard93} are abrupt warmings occurring during glacial periods. They were first observed in the Greenland records, but their signature may be found across the globe \citep{Wolff10aa, Barker11aa, Harrison10aa}, including in the CO$_2$ record \citep{Ahn14aa} (see also Ditlevsen, this volume). 
Heinrich events \index{Heinrich events} are another type of millennial-scale events, defined by the presence of large fractions of ice rafted debris in the North Atlantic, and interpreted as the consequence of icebergs release and melt. Physically, these icebergs induce, when they melt, a negative buoyancy forcing in the ocean that perturbs the ocean circulation. In turn, these circulation changes affect the growth and melt dynamics of ice sheets \citep{Timmermann10ab}. 

There are conceptual models for the dynamics of Dansgaard-Oeschger and Heinrich events \citep{Verbitsky94aa, Paillard95aa, Schulz2002oscillators, Alvarez-Solas10aa, Rial13ab, Kwasniok12aa}. Interestingly, the authors or co-authors of these models often developed conceptual models of ice ages  as well. However, to our knowledge, no fully  coupled ice age / Heinrich / Dansgaard-Oeschger conceptual dynamical system has been published so far, for no obvious reason except perhaps for the fact that it is
 difficult to quantify the effects of an abrupt event on the growth of ice sheets sufficiently realistically.  
\cite{Schulz2002oscillators} and \cite{Sima04aa} modelled explicitly the one-way influence of ice volume on millennium oscillations, but they did not account for the two-way coupling between  ice ages and millennium oscillations.

Besides these abrupt events, high-resolution calcium and isotopic data from ice-cores, and specifically Greenland ice cores, are suggestive of a non-Gaussian nature of fluctuations \citep{Ditlevsen96aa,  Ditlevsen99aa}, with temporal characteristics of intermittency \index{intermittency} and existence of long-range dependence \citep{Marsh97aa, Ashkenazy03aa}.  \index{long memory}
These observations invite us to evaluate additive Gaussian noise parameterisations against more complex alternatives, but we note again that a satisfactory theory must provide a physical rationale for the nature and amplitude of these parameterisations. This is what we are concerned with now. 

\subsection{Parameterising decadal to centennial-variability in EMICS}
Deep-ocean dynamics appear to be the obvious candidate for generating multi-millennial variability in the climate system. One method to determine whether ocean dynamics may generate features such as Dansgaard-Oeschger is to solve ocean geophysical fluid dynamics in more or less idealised configurations, such as idealised geometry and constant or parameterised heat exchanges with the atmosphere.  Some  of these models are simple enough to be analysed formally with concepts and tools of dynamical systems analysis \citep{Dijkstra05aa, Dijkstra05ab,  Colin-de-Verdiere10aa}. There is also a reasonably abundant literature on the generation of millennial ocean variability in higher resolution models, up to GCM level \citep{Vettoretti15aa}. 
However,  this natural ocean variability may also enter in resonance with other sensitive components of the climate system such as ice sheets, and give rise to possibly complex phenomena 
that characterise the glacial periods \citep{Schulz2002oscillators, Schulz2004Glacial-intergl}. 

At time of writing, one of the most detailed simulations of the millennial variability arising from coupled ocean-ice-sheet dynamics was obtained with the CLIMBER model \citep{Ganopolski10aa, Ganopolski12aa}. CLIMBER combines a very simplified ocean model (zonal averages in each basin) with a fairly detailed 3-D ice sheet model. The model follows a deterministic formalism, and the dynamics are complex enough to yield spontaneous episodes of ice discharges influencing the ocean-atmosphere system. Greenhouse-gas concentrations are prescribed in these simulations. 
This model generates a remarkable amount of millennial variability but unfortunately the authors did not provide a log-log plot of the spectrum that would allow us to estimate the spectral slope and the quality of the power-spectrum below the millennial scale. There is however a general argument that high resolution is needed to generate patterns of interannual and interdecadal variability, and a model such as CLIMBER clearly underestimates this variability \citep{petoukhov00}.

We now reach a point of the discussion where it is useful to compare our problems with those faced in weather and seasonal forecasting. 
\citet{Epstein69aa} and \cite{Hasselmann76} early understood the interest of stochastic differential equations to capture the range of climate and weather phenomena, be it for prediction \citep{Epstein69aa} or for the development of a consistent theory \citep{Hasselmann76}. Both objectives---prediction and theoretical consistency---have  stimulated the development of a rich stochastic theory of climate dynamics, reviewed in
\cite{Franzke15aa} and in other chapters of the present book. Among these developments, subgrid-scale stochastic parameterisations \index{stochastic parameterisation} are  increasingly seen as an alternative to deterministic ones. 
The stochastic formalism provides a means to account for the effects of spatio-temporal variability below the resolved scales, on the scales which are effectively resolved by the model. 
\cite{Plant08aa} provides such an example of stochastic parameterisation for cumulus dynamics. 
The principles that have ruled the development of deterministic parameterisations remain valid: a parameterisation must be physically justified, and  improve the model skills. 

Let us then come back to the ice age models.  We mentioned above several examples of stochastic palaeoclimate models, and we also saw that non-Gaussian, or multiplicative processes may be needed to account for the observations. However, we currently lack a way to justify the nature of stochastic parameterisations from our knowledge of internal climate interactions associating ocean, ice sheet dynamics and, possibly, carbon cycle dynamics. 
We suggest that low-resolution models such as EMICs may act as a relay connecting weather dynamics with palaeoclimate dynamics, if we correctly represent the effects of weather and `macro-weather' dynamics in such models using stochastic parameterisations.
The open problem is to develop a methodology to this end. Here, we tentatively propose one.

The Linear Inverse Modelling  (LIM) \index{linear inverse modelling} was introduced as a powerful approach to predict the dynamics of complex spatio-temporal sea-surface temperature patterns such as El-Ni\~no \citep{Penland93aa} \index{El-Ni\~no}.  LIM follows a logic of stochastic dynamical system identification, possibly combined with balance conditions \citep{Penland94aa}, based on observational data. \cite{Kondrashov06ab} further extends the strategy to include non-linear terms, but notes also that data generated by a GCM, rather than observational data, can be used to develop the stochastic model. The model identification procedure is in this case termed  EMR for `Empirical Model Reduction', to indicate the fact that the GCM was reduced to a system of stochastic differential equations following a data-driven process. The formalism was recently generalised by \cite{Kondrashov15aa}. We propose to  include an EMR-type model for interannual and decadal type variability based on GCM output, in a lower-resolution model that will propagate the effects of these  stochastic dynamics up to the millennial scale and beyond. 
A difficulty arises from the fact that we expect this variability may be affected by the background climate conditions. This is clearly the case of El-Ni\~no \citep{Otto-Bliesner99aa}. Hence, we need to express this dependency.

\begin{figure}
\begin{eqnarray*}
\xymatrix{
         &  *+[F.]{\textrm{Simulations at design points only}}  \ar@{=>}[d] & && \\
           \txt{Climate input \\ (ice boundary conditions, CO$_2$)}\ar[r]      \ar@/_3pc/[rr]_{\txt{meta-model (emulator)}}
 & 
    *+[F]        {\textrm{GCM output (time series)}}  \ar[r]& 
           \txt{parameters of stochastic \\ model of mean and variability} \\
} 
\end{eqnarray*}
\caption{\label{fig:metamodel}
Strategy for designing stochastic parameterisations adapted to palaeoclimate modelling. A GCM is used to generate time series as a function of various inputs, including CO$_2$ concentration, ice boundary conditions, astronomical forcing, etc. These time series are used to calibrate a stochastic model for the mean and variability. It may be linear (auto-regressive process) or more complex \citep[e.g.][]{Kondrashov15aa}. A meta-model (emulator) can calibrated to predict stochastic parameterisations valid for any input, on the basis of the experiments performed. }
\end{figure}
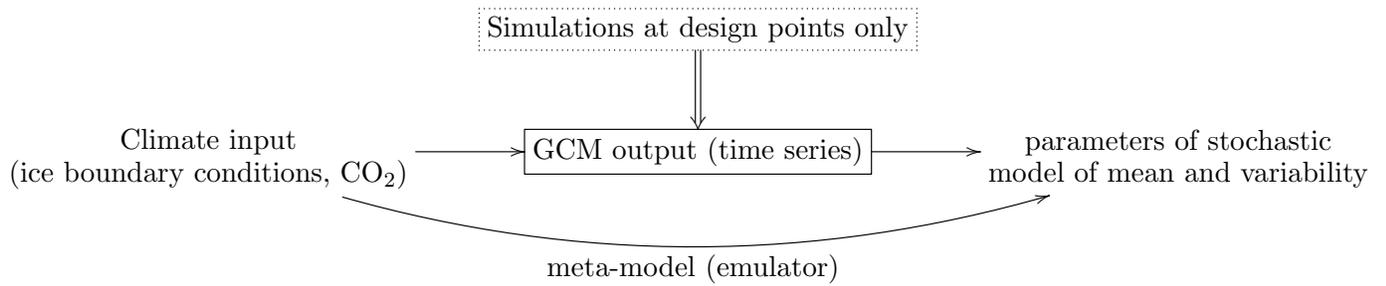

Figure \ref{fig:metamodel} illustrates how this objective can be achieved. 
The  idea is to express relationships between GCMs inputs (inputs are ice volume, CO$_2$ concentration, astronomical forcing) and their output  which characterise (macro-) weather statistics.
This task can be addressed following the theory of computer modelling experiment design and analysis \citep{santner03, Oakley02aa}.  
The principle consists in designing a `meta-model' that estimates the GCM output, based on a set of experiments run with the actual GCM used to calibrate the meta-model. In this context the meta-model is sometimes called an `emulator' \index{emulator}, to emphasise the fact that it is used as a surrogate of the actual GCM.
Palaeoclimate applications along this line include \citet{Araya-Melo15aa}, \cite{Bounceur15aa} and \cite{Holden15aa}. These applications focus on the mean state simulated by the GCM.  They need  to be extended to include aspects associated with the variability. To this end, we can post-process all the GCM experiments of the design by EMR. This  yields, for each experiment, a set of coefficients of a stochastic model. 
If the  emulator is trained on these coefficients, it may then be used to estimate a stochastic model of variability for any input set. We then propose to introduce such stochastic models in Earth-Models of Intermediate Complexity to improve their representation of weather and ``macro-"weather variability \index{macro-weather}, with the objective of analysing the effects of this variability on the slower dynamics of the climate system. 

\subsection{Develop statistical concepts and methods for palaeoclimate dynamics \label{sect:stat}}
Uncertainties in weather and climate modelling arise from different sources: uncertainties in initial conditions, in sub-grid scale processes, in forcing and, last but not least, structural uncertainties pertaining to inadequate representations or simplifications in the climate model. The relative importances of these different sources of uncertainties is partly a subjective matter because they depend on what one wants to model and at what time scale. In general, though, it may be said that structural uncertainty increases with the time scale. Put it differently, we are more uncertain about the nature of the processes which generate the climate dynamics, than about those generating weather dynamics.

These considerations have important consequences on the way we formulate the problem of statistical inference. 
A statistical model generally defines a  \textit{likelihood} function. This is a function of model parameters that is defined once the observations are known. If the framework is Bayesian, the posteriors on uncertain quantities are obtained by multiplying the priors by the likelihood. If we are looking for a point estimate, then, depending on the context, we may simply maximise the likelihood or the posterior. \index{likelihood}

If we adopt the radical hypothesis that the physical  model is perfect, then the shape of the likelihood function reflects only our uncertainties on observations. It is also assumed that the model state contains all the information needed for a one-step-ahead prediction. While  this hypothesis may be effective for weather forecast, it is untenable in climate applications. There is a discrepancy between the physical model and reality. 
As a solution to this problem, \citet{Goldstein04aa} suggest to model reality as the combination of the physical (or mechanical) model, which encodes explicitly our judgements about the interactions and forcing, with a model of the discrepancy. \index{discrepancy} \index{mis-specification}

One simple example will show how this may be achieved in practice (see also \citet{Rougier13aa}).
Suppose that we have a~process model encoded in the form of  deterministic differential equations that determine the value of $x(t)$ given initial conditions $x(0)$ and a~parameter vector $\theta$. 
Symbolically, the equation $\mathrm{d}x = f(x, \theta) \mathrm{d}t$ needs to be integrated forward in time from initial conditions. 
\begin{enumerate}
\item We could transform the equation into a~stochastic increment, for example: $\mathrm{d}x = f(x, \theta) \mathrm{d}t + \sigma \mathrm{d}w$. The stochastic increment is \textit{not} necessarily a representation of sub-grid scale processes. It represents discrepancy in a more general sense. 
This is the approach of \cite{Crucifix09aa}. 
\item We could define a~relationship between the modelled state and the real world, i.e. $y(t) = x(t) + \varepsilon(t)$, where $y$ is the real world, $x$ the modelled state, and $\varepsilon$ a~time-process with its own model, e.g., Orstein-Uhlenbeck process with zero mean. 
\item We could apply some statistical descriptor on the model output in order to extract the only information that we estimate relevant: the mean over time, variability and  higher moments, the spectral peaks, and calibrate the model on the basis of this information. % \hl{ \cite{Xia11aa} call this `feature matching'. }
\end{enumerate}

Out of the three options, option (1.) is closest to the current meteorological practice of stochastic parameterisations.
In the particular example that we gave the judgement is that model discrepancy is effectively captured by uncorrelated Brownian fluctuations that act independently on the system state. 
 An  auto-correlated stochastic process  \citep{Arnold13aa}, or state-dependent discrepancies may be better alternatives. 
 Whichever choice,  the augmented model (mechanics + discrepancy) is then statistically identified as the truth,  which it is not. Some deceiving effects may then occur. For example, the likelihood optimisation procedure may result in compensating for model error by increasing $\sigma$, with possible deleterious effects on the dynamics of the model. \citet{Xia11aa} discusses the origin of the problem. 
Option (3.) follows a radically different strategy. 
The principle is to reduce the amount of information provided by the model, and only match this information with the data. 
As observed by \cite{Wilkinson13aa}, this is nothing but an implicit way of  expressing model discrepancy. 
This approach may  be harder to fit within a~Bayesian formalism because the likelihood may not be available in a tractable form and this question is discussed with possible solutions in \cite{Wilkinson13aa}, \citet{TinaToni02062009} and \citet{Xia11aa}.

We believe that the formulation of discrepancy models for dynamical systems is one area for which conceptual advances are most needed. On the other hand, the algorithmic difficulties pertaining to dynamical system model calibration with uncertain parameters should not  be underestimated. Statistical state-of-the-art methods for dynamical system calibration tend to rely on Monte-Carlo algorithms. They  require multiple or numerous integrations of the dynamical systems and will generally struggle to accommodate more than a dozen parameters \citep{Andrieu10ab, Chopin12aa,  Ionides06aa, Ionides15aa}. In other words, what \cite{Breto09aa} refers to as `time series analysis with mechanical model' \index{time series analysis} can only be attempted with low-order models. It seems risky to venture in predicting what strategies  will prove successful for  tackling the curse of dimensionality and actually do parameter estimation in  GCMs based on palaeoclimate time series, but we imagine that methods of dynamical system identification and spatial reduction such as suggested by \cite{Young11aa}, associated with Gaussian process emulators, can be a part of the solution.

Palaeoclimates applications come with a further source  of difficulty: the observation itself is the end result of a complex and uncertain process of transformation of the climatic signal: formation of the record, preservation, accumulation, etc.  There is a fair amount of statistical literature tackling specifically the issues of signal transformation and chronological uncertainties in the palaeoclimate context \citep{Li10aa,  Parnell14aa,  Werner15aa}. In these articles the inference is made tractable by modelling dynamics with linear auto-regressive systems. Generalisation to non-linear dynamical systems is  a serious challenge \citep{Rougier13aa}. 

\section{Conclusion}
Palaeoclimate dynamics is a~vast field of research. We concentrated here on some mathematical aspects of this research, and identified three areas where state-of-the art mathematics are routinely applied: Dynamical systems theory, numerical simulations, and statistical inference. 

Dynamical systems theory provides concepts to understand the aspects associated with the prediction and characterisation of different modes of variability. 
As we have seen, challenges are associated with the richness of the modes of variability, their multi-scale structure, and a~range of phenomena typical of complex systems such as sensitive dependence on parameters and long-memory \index{long memory}. 

Numerical simulations constitute another pole structuring climate research: they are needed to document the causal 
relationships between local physical processes and emergent phenomena. 
 The complexity and diversity of processes to be accounted for is such that a~compromise needs to be found between what is actually simulated and what is parameterised. Such choices characterise the diversity of models, and a~fuzzy boundary exists between models of intermediate complexity and so-called global climate models. In practice, this  boundary corresponds roughly to the distinction between the objective of simulating the non-stationary \textit{climate regime}, and that of simulating weather and \textit{macro-weather} statistics. 
In this context, we suggested that stochastic parameterisations may constitute a~useful approach to synthesise knowledge obtained from the different model types. Our final goal is to provide a~consistent theory of climate system dynamics spanning several orders of magnitude in the frequency spectrum.  Statistical methods to manage complex simulators, such as meta-modelling and experiment design, provide a~methodological basis to be further explored in the near future.

Statistical inference is the third pole of applied mathematical concepts developed here. Bayesian statistics constitute an attractive framework for inference because it provides a~rationale for parameter and state estimation in presence of limited data~and uncertain structure. The actual potential of  Bayesian approaches merits consideration, but it also comes with pitfalls. 
We noted that discrepancy models may be difficult to characterise, and because of this, the definition of a~likelihood function remains a~subjective process. We add here two points to this discussion.
On the one hand,  joint parameter-state estimation require complex algorithmic procedures that may induce opacity in the scientific process. On the other hand, we must also acknowledge  some difficulties with the validation of probabilistic forecasts in the palaeoclimate context. Contrary to weather forecast, which provide daily opportunities for verification, we cannot claim here for a~frequentist validation of our probability statements. 
The interest of a~Bayesian approach in the palaeoclimate context is  mainly in its being effective for tracing the propagation of uncertainties and identify weak links in our understanding of the climate system. 
 
 We observed that mathematical applications in palaeoclimate research must address numerous difficulties: dynamics are non-autonomous, multi-scale and incompletely characterised; the physical processes are very diverse and some of the important processes may have been overlooked. Finally the data are much less  abundant than in a weather forecast; what they actually record is not always clear and dating is uncertain, too.  
 
 Facing these challenges, it may be tempting to question the relevance of palaeoclimate research programmes, especially in the context of future climate change.
 It is customary to answer this question by referring specifically on the problem of estimating climate sensitivity to CO$_2$ changes.  It is certainly correct to state that palaeoclimate data generally support our current understanding of climate's response to rising CO$_2$ \citep{PALEOSENSE-project-members12aa}, though it is hard to quantify this statement in terms of actual probability distributions. 
 We feel however that it should also be stressed that palaeoclimate data provide the only natural source of information about the slow dynamics of our climate system. They allow us to address the more general problem of the sensitivity of dynamical modes of climate variability. Such information is crucial to assess the long term consequences of climate perturbations. In particular, palaeoclimate data and climate theories have been essential in the debate about the influence of climate dynamics on early humans and civilisations, back to prehistoric and historic \citep{ruddiman07rge}. They are also necessary to forecast the long term anthropogenic effects  on climate \citep{archer05future}. In a sense, ambitious palaeoclimate data recovering programmes such as Antarctic or Arctic drilling programmes can be compared with the Voyager space missions: They make us feel the boundaries of the system we live in.

%\section*{Acknowledgement}
% This research is a contribution to Belgian Federal Policy Office project 'STOCHLIM'. 


\newpage
\clearpage
\bibliography{BibDesk.bib}
\printindex
\newpage

\end{document}